\documentclass[3p,times]{elsarticle}

\usepackage{ecrc}
\usepackage{pstricks}

\volume{00}

\firstpage{1}

\journalname{Nuclear Physics A}

\runauth{Clint Young et al.}

\jid{nupha}

\usepackage{amssymb}
\usepackage{subfigure}
\usepackage{float}

\usepackage[figuresright]{rotating}

\begin{document}

\begin{frontmatter}

\title{Realistic modelling of jets in heavy-ion collisions}
\author{Clint Young}
\address{Department of Physics, McGill University, 3600 University Street, Montreal, Quebec, H3A\,2T8, Canada}
\author{Bj\"orn Schenke}
\address{Physics Department, Bldg. 510A, Brookhaven National Laboratory, Upton, NY 11973, USA}
\author{Sangyong Jeon}
\address{Department of Physics, McGill University, 3600 University Street, Montreal, Quebec, H3A\,2T8, Canada}
\author{Charles Gale}
\address{Department of Physics, McGill University, 3600 University Street, Montreal, Quebec, H3A\,2T8, Canada}

\dochead{}

\begin{abstract}
The reconstruction of jets in heavy-ion collisions provides insight into the dynamics of hard partons in media. Unlike the spectrum of single hadrons, the spectrum of jets is highly sensitive to 
$\hat{q}_{\perp}$, as well as being sensitive to partonic energy loss and radiative processes. We use \textsc{martini}, an event generator, to study how finite-temperature processes at leading order affect dijets.
\end{abstract}

\end{frontmatter}

\section{Jets in a thermal medium}

Dijets are the aftermath of $2\to2$ scattering in perturbative QCD: when the scattering is at a momentum scale significantly above $\Lambda_{{\rm QCD}}$, perturbation theory works well to 
determine the inclusive cross section. The two outgoing partons are born high in virtuality and evolve down through collinear processes; this leads to two narrow cones of partons centered 
on axes determined by the original partons' momenta. Hadronization is non-perturbative and soft, and leads to hadrons centered on these axes. Because the initial partons had relatively 
small total transverse momenta, the total transverse momenta in the two jets should be roughly balanced: ${\bf p}_{T1} \approx -{\bf p}_{T2}$.

This picture leads one to think that the total cross section for dijets can be easily compared with perturbative calculations for inclusive $2\to2$ scattering. However, even in proton-proton collisions with relatively small backgrounds, a method for deciding what particles belong in a jet must be made. A deceivingly simple method is to use the anti-$k_T$ algorithm 
\cite{Cacciari:2008gp}:
a metric defines a ``distance" between two 4-momenta; for infrared and collinear $1 \to 2$ processes, this metric is defined so that the distance between the two final momenta are small. A list of 
distances between all pairs of momenta in a given event is then determined and ordered from smallest to largest, and the list is shrunken by combining these momenta, starting at the beginning of 
the list and stopping once all distances are above some lower limit. The final momenta, when they are sufficiently large, can be called reconstructed jets. Because the shortest distances are added first, the algorithm is insensitive to infrared and collinear physics.

The large cross sections for jet production, as well as the large energy separation between jet energies and the highest temperatures reached, allowed jets to be one of the first heavy-ion observables examined at the Large Hadron Collider \cite{Aad:2010bu, CMSdijets2011}. A large sample of dijets were reconstructed from lead-lead events measured at ATLAS, and the distributions in dijet asymmetry $A_j \equiv (E_>-E_<)/(E_>+E_<)$, where $E_>$ ($E_<$) is the energy of the (second-)largest jet, were determined for several centrality classes. In the 0-10\% centrality class, the distribution $dN/dA_j$ was determined to be flat from $A_j = 0$ up to $A_j = 0.5$, significantly different from all preceding proton-proton analyses where the distribution is strongly peaked at $A_j = 0$. 

\begin{figure}
\centering
\mbox{\subfigure{\includegraphics[width=3in]{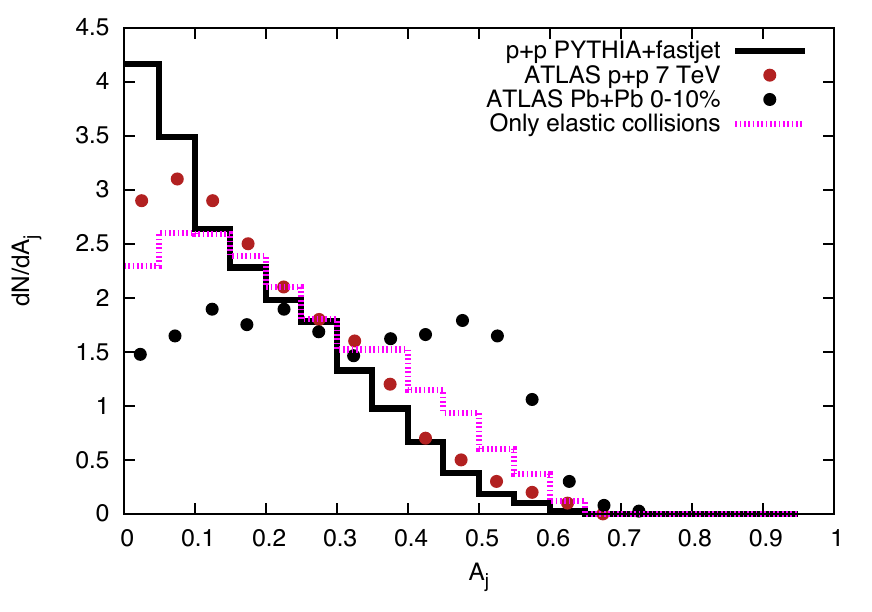}}
\quad
\subfigure{\includegraphics[width=3in]{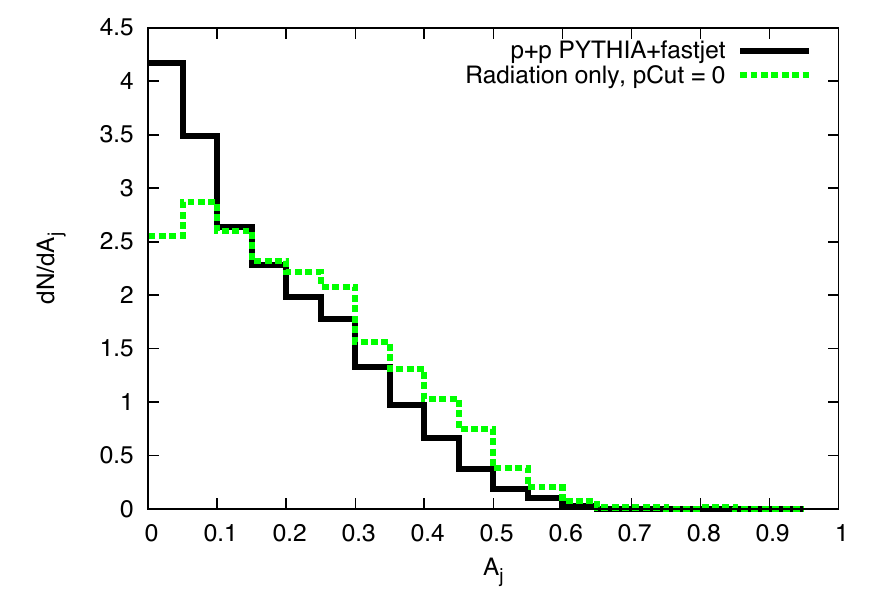} }}
\caption{(Colour online) The effect of elastic and radiative processes on dijet asymmetry. Left: dijet asymmetry of proton-proton collisions (black) at 2.76 TeV compared with the $A_j$ distribution 
in 0-10\% lead-lead collisions considering {\it only elastic processes}. Data from ATLAS for p+p and 0-10\% Pb+Pb collisions are shown, with (red and black) solid circles. Right: the dijet asymmetry in 0-10\% lead-lead collisions considering {\it only radiative processes}. } 
\label{elasticVsRadiative}
\end{figure}

We consider the propagation of the partons of a jet through a deconfining, thermal medium. Determining the propagation perturbatively leads first to considering elastic $2 \to 2$ scattering at tree-level, and using the hard thermal loop (HTL) approximation leads to the rate 
\begin{equation}
\frac{d\Gamma(p, p^{\prime})}{dp^{\prime  3}} = \frac{d_k}{16p^0 p^{\prime 0} } \int \frac{d^3k}{(2\pi)^3 \omega_k} \frac{d^3k^{\prime}}{(2\pi)^3 \omega_{k^{\prime}} } | {\cal M} |^2
(2\pi)^4 \delta^4(p+k-p^{\prime}-k^{\prime}) n(\omega_k)[1 \pm n(\omega_{k^{\prime}})] {\rm .} \nonumber
\end{equation}
When interference between multiple elastic scatters is small, these rates can be convoluted with a distribution of partons and calculated in the rest frames of the medium and at temperatures 
determined by a model for the heavy-ion collision. Figure \ref{elasticVsRadiative} shows how convoluting elastic rates can lead to some dijet asymmetry. Using $\alpha_s \approx 0.3$, and only elastic processes, does not lead to a large average dijet asymmetry.

However, elastic scattering is not the only process contributing to the evolution of jets at leading order in $g$. At high energies, $2 \to n$ scatterings where $n>2$ become collinearly enhanced. 
All that keeps these diagrams from being singular in the collinear limit is the thermal mass squared $\sim g^2T^2$, which causes scatters with gluon emission to contribute at a smaller order in 
$g$ than one might na\"ively expect. However, at the high temperatures where HTL effective theory applies, the medium is also very dense, meaning that the Landau-Pomeranchuk-Migdal (LPM) effect must also be taken into account. Multiple scatterings can be resummed using an integral equation \cite{Arnold:2002ja}. Figure \ref{elasticVsRadiative} also shows the effect of radiative splittings alone. These radiative splittings, which are in addition to the splittings that occur in the evolution from high to low virtuality, lead to a distribution in dijet asymmetry almost the same as the distribution predicted for proton-proton collisions at this energy. This should not be a surprise: the anti-$k_T$ algorithm is designed to be insensitive to collinear processes.

However, what happens when both elastic and radiative processes are taken into account? Radiation evolves the distribution of partons in a jet down to small momentum fractions. With elastic scatterings occurring, the physics of $\hat{q}$ also is taken into account. Given a parton of momentum $p$, an elastic scattering transferring transverse momentum $\sim gT$ will deflect the parton by $\delta \theta \sim gT/p$. Thanks to the evolution down to small momentum fractions, many more of these partons will have small $p$ and will be deflected more. This identifies a possible source for dijet asymmetry in the absence of strong energy loss: enhanced decollimation of the jet caused by in-medium jet evolution.

\section{The importance of full Monte Carlo}

The evolution of a distribution of partons in a thermal medium can be described with this integral equation:
\begin{eqnarray*}
\frac{dP_{q\bar{q}}(p)}{dt} &=& \int_k P_{q\bar{q}}(p+k) \frac{d\Gamma^q_{qg}(p+k,k)}{dkdt} -P_{q\bar{q}}(p)\frac{d\Gamma^q_{qg}(p,k)}{dkdt} \\
& & +P_g(p+k)
\frac{d\Gamma^g_{q\bar{q}}(p+k,k)}{dkdt}{\rm ,} \\
\frac{dP_g(p)}{dt} &=& \int_k P_{q\bar{q}}(p+k)\frac{d\Gamma^q_{qg}(p+k,p)}{dkdt}+P_g(p+k)\frac{d\Gamma^g_{gg}(p+k,k)}{dkdt} \\
 & & -P_g(p)
\left( \frac{d\Gamma^g_{q\bar{q}}(p,k)}{dkdt}+\frac{d\Gamma^g_{gg}(p,k)}{dkdt}\Theta(2k-p) \right){\rm ,} \\
\end{eqnarray*}

If these rates are small unless $k$ is small compared with $p$, then one might want to make a ``diffusive approximation" that turns these equations into differential equations. However, for massless partons, this was demonstrated in Figure 3 in \cite{Schenke:2009ik} to pose difficulties: the initial distribution $\delta(E-10\;{\rm GeV})$ was evolved for 5 fm/c according both to integral equations and the Fokker-Planck equation. While both approaches determined similar mean values, the Fokker-Planck equation predicted a significantly smaller distribution at small $E$ than solving the integral equation did.

Our description of jet modification depends critically on determining the jet shape properly at small momentum fractions; it is clear that the full integral equation must be solved. However, solving these equations in a rapidly evolving and expanding thermal medium is prohibitive and indeed, undesirable. Monte Carlo simulations are ideal for solving these integral equations: they are not significantly complicated by the addition of new processes like other numerical solutions, and they are adaptable.

\section{Dijet observables at RHIC and the LHC}

In \cite{Young:2011qx}, we showed how dijet asymmetry might be described as the result of parton evolution at finite temperature, using \textsc{martini}, an event generator for high-momentum observables in heavy-ion collisions \cite{Schenke:2009gb}. With it, jets were sampled using \textsc{pythia8} \cite{Sjostrand:2006za}, and evolved according to finite-temperature rates using the results of \textsc{music}, a solver of the 3+1-dimensional relativistic hydrodynamical equations which describe well the bulk observables at RHIC and the LHC \cite{Schenke:2010nt}. After hadronization, jets are reconstructed using \textsc{fastjet} \cite{Cacciari:2006sm}, to be consistent with the techniques used by the experimentalists.

\begin{figure}
\centering
\mbox{\subfigure{\includegraphics[width=3in]{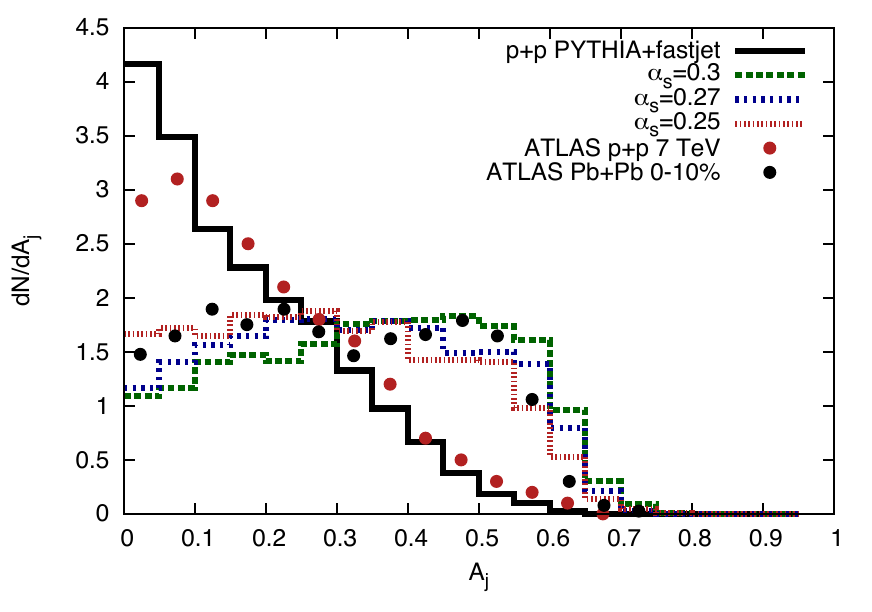}}
\quad
\subfigure{\includegraphics[width=3in]{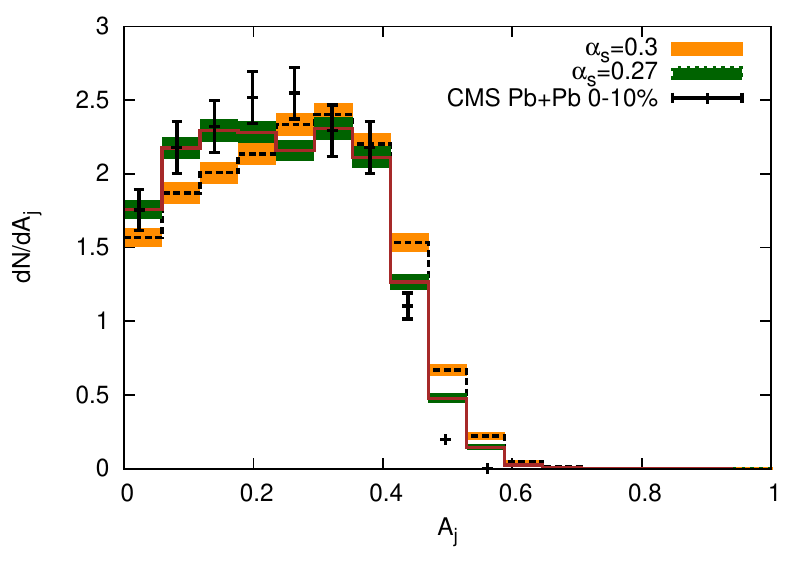} }}
\caption{(Colour online) $dN/dA_j$, including both elastic and radiative processes, with experimental cuts, compared with dat from \textsc{ATLAS} and \textsc{CMS}. } 
\label{allProcesses}
\end{figure}

The main results of our work on dijets are shown in Figure \ref{allProcesses}. Applying both the elastic and radiative processes to the partons in a jet leads to agreement with the results of both 
ATLAS and CMS, using $\alpha_s \sim 0.3$, unlike the results found using elastic or radiative processes alone.

\section{$R_{AA}$ and the effect of running coupling and finite-size effects}

\begin{figure}[H]
\centering
	\includegraphics[trim = 0mm 0mm 0mm 0mm, clip,width=3in]{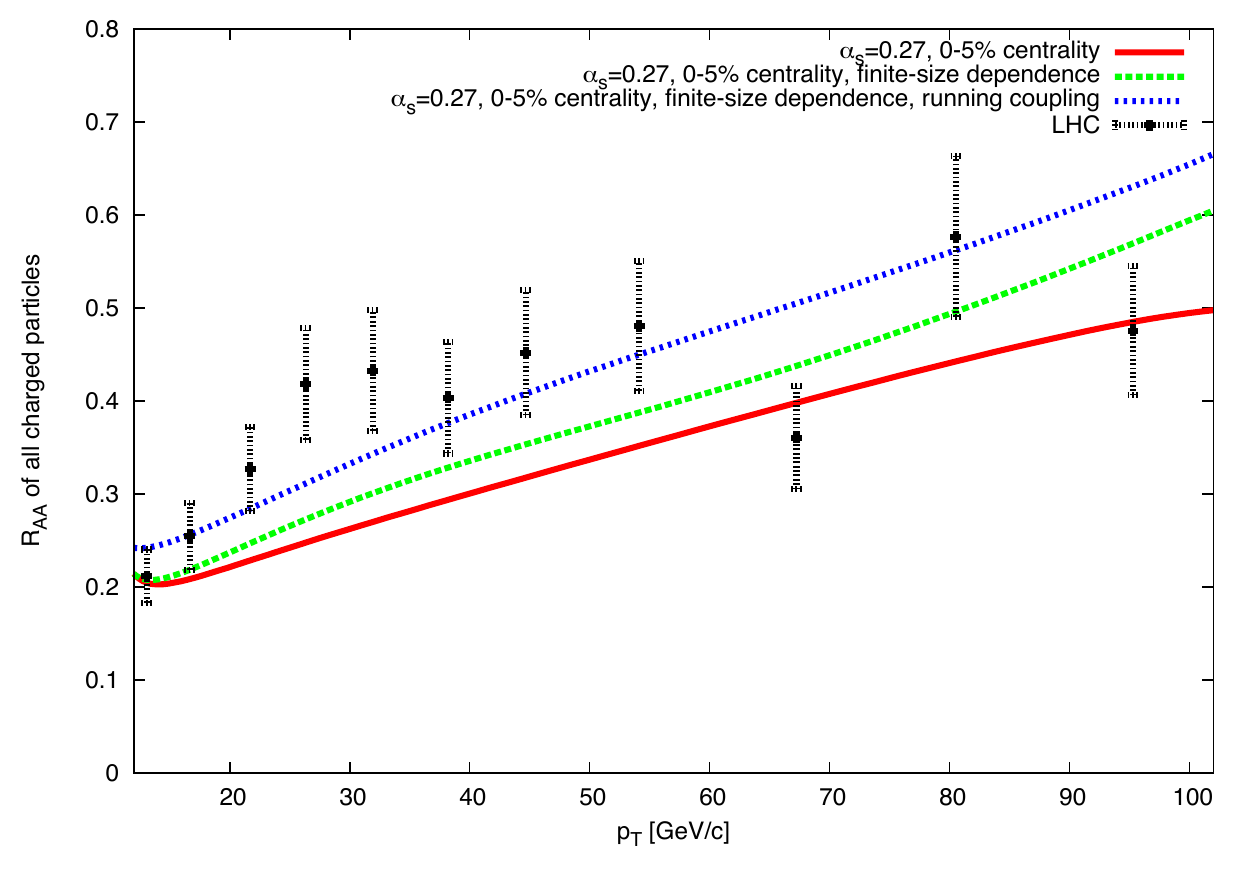}
	\caption{(Colour online) The nuclear modification factor $R_{PbPb}(p_T)$, including the effect of finite-size and running coupling, compared with preliminary CMS results.}
	\label{RAA}
\end{figure}

Finally, we present preliminary results for the charged particle nuclear modification factor $R_{PbPb}(p_T)$ in the 0-5\% centrality class at the LHC, in Figure \ref{RAA}. Three curves are compared with data: first, the results using the same radiative rates as was used to describe dijet asymmetry are plotted with the solid (red) curve. A slow rise in $R_{AA}$ with $p_T$ is observed, consistent with the LPM effect \cite{Turbide:2005fk}. Next, the dashed (green) curve shows results with a parametrization of finite-size effects. For a medium with finite size, the interference between vacuum and medium-induced splittings suppresses the radiative rates at early times \cite{CaronHuot:2010bp}. Finally, the dotted (blue) curve shows the results taking account of both finite size effects and the running of the coupling. The momentum scale of the splitting vertex is given by $\left \langle k_T^2 \right \rangle = \hat{q}t_f = \hat{q} k/\left \langle k_T^2 \right \rangle$, leading to $\left \langle |k_T| \right \rangle = (\hat{q}k)^{\frac{1}{4}}$. Momentum transfers in elastic collisions are space-like and of order $gT$ and their running is ignored.

At this point, both effects are applied approximately to total radiative rates and not to the differential rates. To include the effect of running coupling without this approximation is simply a matter of 
recalculating the rates. However, exact implementation of finite-size effects poses a challenge to Monte Carlo simulations, because they depend on the previous positions of the partons and the evolution of the medium. Perhaps using the initial position of a given parton, and the temperature of the medium at this point, can be used to estimate the temperature profile for the parton's evolution with enough accuracy that the finite-size effect can be taken into account accurately.

\end{document}